\documentclass[useAMS,usenatbib]{mn2e}
\usepackage{natbib}
\usepackage{epsfig}
\usepackage{amsbsy}
\usepackage{hyperref}
\usepackage{ amssymb }

\providecommand{\adsurl}[1]{\href{#1}{ADS}}

\newcommand{\lya}{Lyman-$\alpha$~}

\newcommand{\eg}{{\it e.g.~}}

\newcommand{\be}{\begin{equation}}
\newcommand{\ee}{\end{equation}}
\newcommand{\ba}{\begin{eqnarray}}
\newcommand{\ea}{\end{eqnarray}}
\newcommand{\brr}{\begin{array}}

\newcommand{\err}{\end{array}}
\newcommand{\bc}{\begin{center}}
\newcommand{\ec}{\end{center}}

\DeclareMathAlphabet{\mathsc}{OT1}{cmr}{m}{sc}
\def\testbx{bx}%
\DeclareRobustCommand{\ion}[2]{%
\relax\ifmmode
\ifx\testbx\f@series
{\mathbf{#1\,\mathsc{#2}}}\else
{\mathrm{#1\,\mathsc{#2}}}\fi
\else\textup{#1\,{\mdseries\textsc{#2}}}%
\fi}
\title[IGM and Coupled Dark Energy]
{The Impact of Coupled Dark Energy Cosmologies on the High-Redshift Intergalactic Medium}

\author[M. Baldi \& M. Viel] 
{Marco Baldi$^{1,2}$ \&
Matteo  Viel$^{3,4}$
\\
$^1$ Excellence Cluster Universe, Boltzmannstr. 2, D-85748 Garching, Germany (marco.baldi@universe-cluster.de)\\
$^2$ University Observatory, Ludwig-Maximillians University Munich, Scheinerstr. 1, D-81679 Munich, Germany\\
$^3$ INAF - Osservatorio Astronomico di Trieste, Via G.B. Tiepolo 11,
I-34131 Trieste, Italy
(viel@oats.inaf.it)\\
$^4$ INFN/National Institute for Nuclear Physics, Via Valerio 2,
I-34127 Trieste, Italy\\
\\}

\begin{document}
\maketitle
\begin{abstract}
We present an analysis of high-resolution hydrodynamical N-body
simulations of coupled dark energy cosmologies which focusses on the
statistical properties of the transmitted \lya flux in the high-redshift 
intergalactic medium (IGM). In these models the growth of the
diffuse cosmic web differs from the standard $\Lambda$CDM
case: the density distribution is skewed towards underdense regions
and the matter power spectra are typically larger (in a scale
dependent way).  These differences are also appreciable in the \lya
flux and are larger than 5 \% (10\%) at $z=2-4$ in the flux
probability distribution function (pdf) for high transmissivity
regions and for values of the coupling parameter $\beta=0.08$
($\beta=0.2$).  The flux power spectrum is also affected at the $\sim
2\%$ ($\sim 5-10\%$) level for $\beta=0.08$ ($\beta=0.2$) in a
redshift dependent way. We infer the behaviour of flux pdf and flux
power for a reasonable range of couplings and present constraints
using present high and low resolution data sets. We find an upper limit 
$\beta \lesssim 0.15$ (at $2\sigma$ confidence level), which is obtained
using only IGM data and is competitive with those inferred
from other large scale structure probes.
\end{abstract}


\section{Introduction}

The observational evidence of the present accelerated cosmic expansion
represents one of the major challenges to our understanding of
the Universe. The standard $\Lambda $CDM cosmological model identifies
the origin of this acceleration with a cosmological constant term in
the field equations of General Relativity.  However, this
interpretation suffers of extremely severe fine-tuning problems, and
possible alternative explanations of the accelerated expansion in
terms of a Dark Energy (DE) dynamical field have been proposed
\citep[as \eg by ][]{Wetterich_1988,Ratra_Peebles_1988,
  ArmendarizPicon_etal_2000}. Among these, particular attention has
been recently devoted to coupled DE models (cDE)
\citep{Wetterich_1995,Amendola_2000,Farrar_Peebles_2004,Baldi_2010}
where a direct interaction between the DE field and Cold Dark Matter
(CDM) particles determines peculiar features in the background
expansion of the Universe \citep[\eg][]{Amendola_2000}, in the
evolution of linear density perturbations
\citep[\eg][]{DiPorto_Amendola_2008}, and even in the nonlinear
dynamics of collapsed structures at small scales
\citep[\eg][]{Baldi_etal_2010,Baldi_2010}.  It is therefore of crucial
importance in the present cosmological investigation to devise
observational tests capable to distinguish between the standard
$\Lambda $CDM cosmology and alternative DE models as \eg the cDE
scenario \citep{Honorez_etal_2010,Baldi_Pettorino_2010}.  In the
present work, we explore the possibility to use the observed
properties of the diffuse baryonic matter at high redshifts as a
direct probe to test and constrain cDE cosmologies.

Standard cosmological models based on cold dark matter plus a
cosmological constant predict that most of the baryons at high
redshift are in a diffuse form, the Intergalactic Medium (IGM), and
fill a significant portion of the Universe, giving rise to the
so-called cosmic web: a network of median fluctuated filaments
interconnecting galaxies and tracing the underlying dark matter
distribution.  A great progress in the study of the IGM has been
recently made thanks to the large data sets available and in
particular high resolution quasar (QSO) spectra ({\em Ultra Violet Echelle
Spectrograph} or HIRES) and the low resolution {\em Sloan Digital Sky Survey}
(SDSS) QSO spectra: the present limitations appear to be of systematic
nature rather than statistical. High and low-resolution \lya QSO
spectra of distant sources are thus very useful in characterizing the
properties of the underlying mass density field at $z=2-6$ along the
line-of-sight \citep[\eg][]{bi,croft02,viel04,mcdonald05,meiksin09}
and are now routinely analyzed to reconstruct the matter distribution
in three-dimensions. The dynamical information on the growth of
structures is however convolved non-linearly with other physical
effects determining the shape of the \lya absorption lines in redshift
space (thermal history, peculiar velocities, etc.), and also
observational procedures (continuum fitting, removing of the metal
lines and strong absorption systems, etc.) need to be properly modeled
and marginalized over \citep[see \eg][]{mcdonald06}. All these issues
make it quite difficult to use the IGM as a cosmological tool since
its structures have to be modeled with hydrodynamical simulations
that incorporate the most relevant physical processes. Nevertheless,
IGM \lya data in combination with other probes provide the tightest
constraints to date on the neutrino mass fraction and on the coldness
of CDM and suggest a higher r.m.s. value for the amplitude of the
matter power than that obtained from cosmic microwave background data
\citep[\eg][]{seljak06,viel08}.

In this work we perform the first high-resolution hydrodynamical
simulations of cDE models with gas cooling and star formation in order
to check whether one can use the IGM to detect the different growth of
cosmic structures predicted by these models compared to the standard
$\Lambda$CDM scenario.  To do so, we make use of the modified version
of the parallel hydrodynamical N-body code {\small GADGET-2}
\citep{springel} specifically devised to include the effects of cDE
models and already presented in \citet{Baldi_etal_2010}. 

The layout of this Letter is as follows:
in Sec.~2 we describe the cosmological models investigated and the
hydrodynamical simulations, in Sec.~3 we present the results in terms
of 1-pt and 2-pt matter and flux statistics, in Sec.~4 we compute the
constraints on the DE-CDM coupling parameter $\beta$ using the SDSS
flux power and the flux pdf of UVES spectra and in Sec.~5 we draw our
conclusions.

\section{hydrodynamical simulations of coupled dark energy models}
We consider cDE cosmologies where the accelerated expansion of the
Universe is driven by a DE scalar field $\phi $ which interacts with
the CDM fluid by directly exchanging energy according to the
equations:
\begin{eqnarray}
\label{continuity1}
\rho '_{c}+3{\cal H}\rho _{c} &=& -\beta \phi ' \rho _{c} \\
\rho '_{\phi }+3{\cal H}\rho _{\phi } &=& +\beta\phi '\rho _{c}\,,
\label{continuity2}
\end{eqnarray}
where a prime denotes a derivative w.r.t. conformal time and ${\cal H}$ is the conformal Hubble function.
The constant parameter $\beta $ fully specifies the interaction and determines the strength 
of the DE-CDM coupling\footnote{Note that we have used units in which $M_{Pl}=(8\pi G)^{-1/2} = 1$ and that 
the definition of the coupling $\beta $ differs by a factor $\sqrt{3/2}$ from the one used in some of the literature.}.
The cDE models described by Eqs.~(\ref{continuity1},\ref{continuity2}) have been widely studied in the literature
\citep[see \eg][and references therein]{Wetterich_1995,Amendola_2000,Pettorino_Baccigalupi_2008,
Wintergerst_Pettorino_2010,Baldi_etal_2010}
to which we refer the reader for an extensive discussion of the main features of these models. For the analysis
carried out in this Letter, the effect of primary interest is the enhanced growth of CDM density perturbations
arising in cDE cosmologies due both to the long-range fifth-force acting between  CDM coupled particles (which
attract each other with an effective gravitational constant $\tilde{G}=G_{N}[1+2\beta ^{2}]$, where $G_{N}$
is the standard Newtonian value) and to the additional velocity-dependent acceleration of coupled particles
$\vec{a}_{v}\propto \beta \dot{\phi} \vec{v}$ which follows from momentum conservation in cDE models.

For our analysis we rely on simulations run with the modified version
by \citet{Baldi_etal_2010} of the parallel hydrodynamical N-body code
{\small {GADGET-2}} based on the conservative `entropy-formulation' of
{\em Smoothed Particle Hydrodynamics} (SPH) \citep{springel}.  They
consist of a cosmological volume with periodic boundary conditions
filled with an equal number of dark matter and gas particles.
Radiative cooling and heating processes were followed for a primordial
mix of hydrogen and helium.  
The star formation criterion simply
converts in collisionless stars all the gas particles whose
temperature falls below $10^5$ K and whose density contrast is larger
than 1000 (\citet{viel04} have shown that the star formation criterion has a
negligible impact on the flux statistics considered here).  More
details on the gas cooling and on the Ultra Violet background can be found in \citet{viel04}.

The cosmological reference model corresponds to a `fiducial'
$\Lambda$CDM Universe with parameters, at $z=0$, $\Omega_{\rm m
}=0.26,\ \Omega_{\rm \Lambda}=0.74,\ \Omega_{\rm b }=0.044$, $n_{\rm
  s}=0.963$, $H_0 = 72$ km s$^{-1}$ Mpc$^{-1}$ and
$\sigma_8=0.796$, consistent with the results of WMAP 5-years data \citep[][]{wmap5}.  
The cDE models considered in our analysis are two
  scalar field models with a Ratra-Peebles \citep{Ratra_Peebles_1988}
  self-interaction potential, $U(\phi ) = U_{0} \phi ^{-\alpha }$,
  with $\alpha =0.143$ and with couplings $\beta = 0.08$ and $\beta =
  0.2$, already studied in \citet{Maccio_etal_2004} and
  \citet{Baldi_etal_2010}, and labelled ``RP2" and ``RP5",
    respectively.  These acronyms refer to the type of self-interaction potential (where ``RP" stands for the Ratra-Peebles potential)
    and to the strength of the coupling in a scale ranging from $\beta = 0.04$ (which would appear as RP1) to $\beta = 0.2$ (our RP5 model). 
    Other numerical simulations for some related models
    of interacting DE or modified gravity -- but without hydrodynamics --
    have been presented by \eg
   \citet{Nusser_Gubser_Peebles_2005,Kesden_Kamionkowski_2006,Schmidt_2009,Keselman10,Li_Barrow_2010}.

Initial conditions for all the runs are realized by displacing
particles from a cartesian grid according to Zel'dovich approximation
in order to obtain a particle distribution with the desired spectrum
of density fluctuations. The initial shape of the power spectrum for
cDE models is slightly blue-tilted with respect to the fiducial
$\Lambda $CDM scenario \citep[see][for a description of the tilt in
  cDE models]{Baldi_etal_2010} and this effect is taken into account
in our runs. The amplitude of the initial ($z=60$) density
fluctuations in the different simulations is then rescaled with the
appropriate growth factor in order to normalize all the linear power
spectra to the same $\sigma _{8}$ at $z=0$.
\begin{figure*}
\includegraphics[width=17cm,height=7cm]{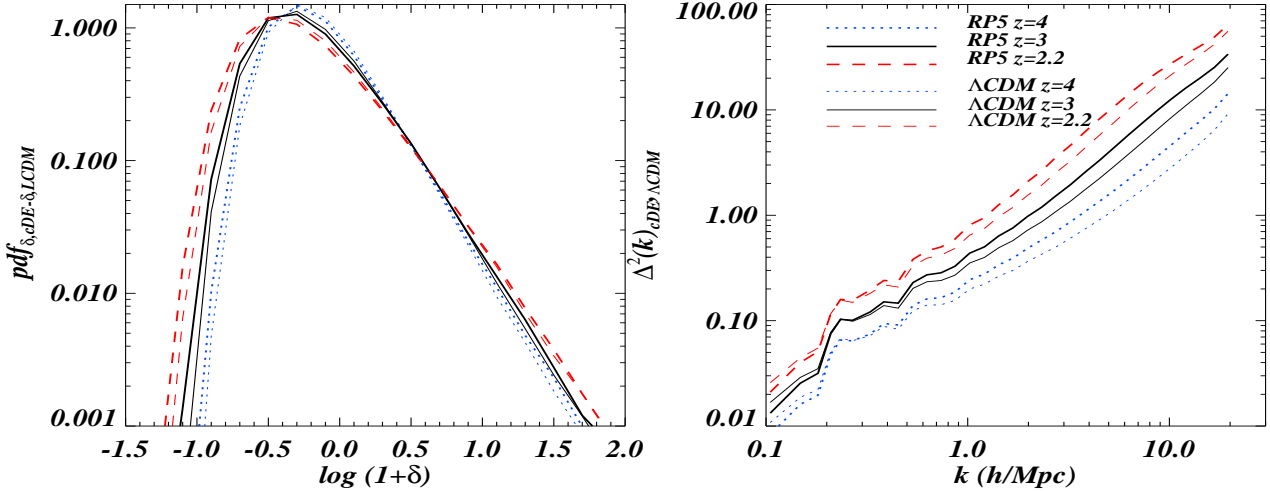}
\caption{{\it Left:} The IGM density pdf of the RP5
  (thick curves), and the $\Lambda$CDM (thin curves) models at
  $z=2.2,3,4.2$ (red dashed, black continuous, blue dotted curves,
  respectively). {\it Right}: Matter power spectra of the $\Lambda$CDM and the RP5 models.}
\label{figdelta}
\end{figure*}
It is also important to stress that the cDE and $\Lambda$CDM
simulations have been performed with the same random phases in the
initial conditions and the same set of cosmological parameters at
$z=0$: thereby the differences that we will highlight will be
exclusively due to the different couplings and not to other
parameters. We have used $2\times 400^3$ dark matter and gas particles
in a $60\ h^{-1}$ comoving Mpc box for the flux power to sample the
scales probed by the SDSS. The gravitational softening was set to
$5~h^{-1}$ kpc in comoving units and the mass per DM particle is $2
\times10^8 $M$_{\odot}/h$.  The snapshots analysed are in the redshift
range $z=2.2 - 4.2$ since these are the redshifts at which the flux
power is measured by SDSS and 1000 quasar spectra are extracted for
each redshift bin. Noise properties are added to the spectra in order
to reproduce the observed data.

\section{Results}
\subsection{Impact on density and flux statistics}

We first consider the matter fields as extracted from the simulations.
In Figure \ref{figdelta} (left panel) we show results for
the IGM density probability distribution functions between cDE models
and $\Lambda$CDM at $z=2.2,3,4$. The red, blue and black curves refer
to the three different redshifts ($z=2.2,3,4$, respectively) while the
thick ones represent the results for RP5 ($\beta=0.2$) and the thin
ones for $\Lambda $CDM (for clarity we do not plot the RP2 model in these figures). 
It is clear that it is more likely to
encounter low density regions along the line-of-sight in a cDE model
rather than in a standard cosmology: i.e. the gas distribution of a
cDE model is skewed towards regions that are less dense than the
mean. While the differences at $z=4$ for the RP2 model are about a
factor two for $\rho~/<\rho>=0.1$, they rise to a factor seven for the
RP5 model. A visual inspection of some one-dimensional gas density
fields confirms this trend and shows that in cDE
models the underdense regions are usually emptier than in the
corresponding $\Lambda$CDM case.  There are also some differences at
large densities but we do not focus on those here since their volume
filling factor is much smaller than less dense regions.

\begin{figure*}
\includegraphics[width=17cm, height=7cm]{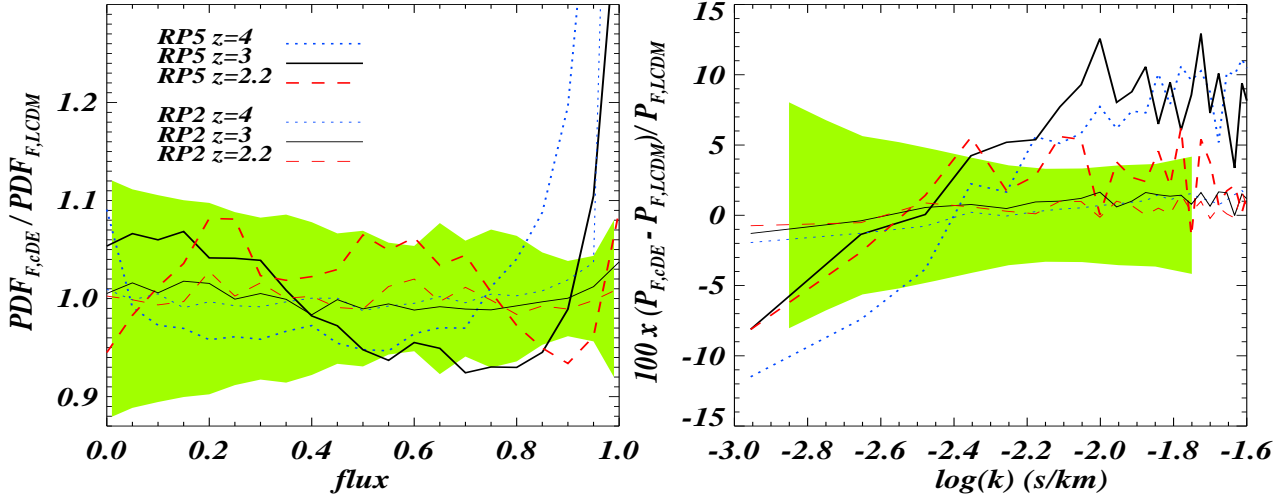}
\caption{{\it Left:} Ratio between the flux pdf of the RP5 (thick
  curves), RP2 (thin curves) and the $\Lambda$CDM models at
  $z=2.2,3,4.2$ (red dashed, black continuous, blue dotted curves,
  respectively). {\it Right:} Percentage differences between the flux
  power of the RP2 and RP5 models and that of $\Lambda$CDM. The shaded
  areas in both panels represent the statistical error at $z=2.94$ for
  the high resolution PDF data of Kim et al. (2007) and for the $z=3$
  bin of the SDSS QSO flux power as computed by McDonald et
  al. (2006). }
\label{figflux}
\end{figure*}

In the right panel we plot the matter power spectra for the same two
models shown in the left panel: the non-linear power spectrum of cDE models is tilted and usually larger than
for the $\Lambda$CDM model. We recall that the power spectra are
normalized to reproduce the same $\sigma_8$ at $z=0$ and thereby their
growth is different \citep[see][]{Baldi_etal_2010}. The pivot scale is
at about $0.25~ h/$Mpc, which corresponds to (roughly) a scale of
about 0.0025 s/km in redshift space for the corresponding one
dimensional flux power.  At scales $1-2\, h/$Mpc the
matter power for RP5 (RP2) is about 25\% (2\%) larger than the
$\Lambda$CDM matter power (we have again plotted here only the $\Lambda $CDM and the RP5 models for clarity reasons).

We note that the inferred linear matter power spectrum amplitude from
the analysis of the SDSS flux power by \cite{mcdonald05} is recovered
with a $1\sigma$ error bar of about 15\%, thus we naively expect, even
without performing any analysis in terms of flux power, that this $k-$
dependent increase in the matter power can be constrained and the RP5
model ruled out from this data set alone.

We now turn our attention to the mock sets of QSO spectra that have
been extracted from each snapshot and normalized to reproduce the
same (observed) effective optical depth $\tau_{\rm eff}(z)=-\ln <F(z)>$ as
estimated by \cite{tkim}.  The scaling factors that are used in order
to normalize the optical depth differ by less than 4\% between cDE and
$\Lambda$CDM models.

In Figure \ref{figflux} we plot the 1-pt and 2-pt distribution
functions in terms of flux pdf (left panel) and flux power (right
panel).  The color coding is the same as in the previous figures.  The
differences that are present for the low-density tail of the gas pdf
show up in regions of high transmissivity (flux values $\sim$ 1). This
demonstrates that cDE voids are emptier of gas and contain less
neutral hydrogen than in a corresponding $\Lambda$CDM universe.  Our
findings therefore suggest that cDE models might alleviate the
discrepancy between simulations and observations in voids, thereby
providing a possible solution to the so called ``Void Problem"
\citep[][]{Peebles_01,Keselman10}. 
It is also intriguing that 
from the analysis of SDSS data \citet{mcdonald05} find a structure growth at $z=3$
that is faster than that predicted by $\Lambda$CDM, although with a relatively low statistical significance:
this could be better reproduced by cDE rather than standard cosmologies.

We also note that there is an increase in regions of low
transmissivity, although this effect is subdominant compared to the
first one: the flux pdf is clearly more skewed in cDE cosmologies than
in the standard case since the cosmic web appears to be more evolved
in the former case rather than in the latter.  The shaded area
represents the $\pm 1\sigma$ statistical error bars inferred from a
jack-knife estimate of high-resolution UVES data \citep{tkim} at
$z=2.94$.

The flux power is shown in the right panel and we can see that a
similar (although smaller in magnitude) trend to the matter power
spectrum case is present. The flux power is tilted with a pivot scale
roughly corresponding to that of the matter power. The differences
consist in a suppression at the 10\% level at the largest scales for
RP5 ($\beta=0.2$) and 2\% for RP2 ($\beta=0.08$): the effect is nearly
symmetric w.r.t. the pivot scale and at 0.01\, s/km the power has
increased by roughly the same amount.  The shaded area indicates for
quantitative comparison the $\pm 1\sigma$ statistical error estimated
at $z=3$ by \cite{mcdonald06} from the low resolution SDSS QSO data
set

If we compare the effects that cDE cosmologies have in terms of flux
pdf and flux power with those of other parameters we can note the
following: $i)$ as for flux pdf we can compare our findings with Figs.
1 and 2 of \cite{bolton07}: the effect is similar to that of having a
larger value for $\sigma_8$ and a lower value for the parameter
$\gamma$ for the IGM temperature-density relation, but the redshift
dependence is quite different in the two cases since we have a strong
evolution between $z=2$ and $z=3$ in cDE cosmologies and a much smaller
one in $\Lambda$CDM; $ii)$ the effect on the flux power is distinct
from that of having a different $\sigma_8$ value for the power
spectrum amplitude (see Fig.~13 in \cite{mcdonald05} and Fig.~3
in \cite{vielhaehnelt06}); $iii)$ the effect on the flux power is also different
from that of a different spectral index which does not allow for a
change of sign but is either positive or negative in the whole range
of scales \citep[Fig.~3 in][]{vielhaehnelt06}. 
To give a rough quantitative reference for a direct comparison
we note that the effect in terms of flux
power of a non-zero coupling, for the particular models investigated
here, is similar to that introduced by a change in the slope of the
linear matter power spectrum \citep[see][]{mcdonald05} 
of about $\Delta n = 0.05$,  $n$ being the slope of the power spectrum at $k = 0.009$ s/km and $z = 3$, 
while the trends are different for the other
cosmological and astrophysical parameters. As for the flux pdf the
effects are similar to those introduced by a different $\sigma _{8}$
value: for fluxes of the order F=0.9 the effect on the flux pdf is
around $\sim 10 \%$ when the $\sigma _{8} $ value is increased by $13\%$,
very similar to those introduced by the RP5 model but with a different
redshift dependence.
We also note that the signature is
degenerate with that of non-gaussianity at the level of flux pdf but
not at the level of flux power \citep{vielng}. Therefore the cDE signature on \lya
flux statistics for the specific models investigated here seems to
be unique and competitive constraints can be expected.

\subsection{Constraints on the coupling}
We now seek constraints on the coupling parameter $\beta$ by using a
second order Taylor expansion method for the flux pdf and flux
power. This method is described in \cite{vielhaehnelt06} and in \cite{lidz09}
and although having the drawback of underestimating the error bars it has
the advantage of being calibrated on accurate full hydrodynamical
simulations.  We perform a Monte Carlo Markov Chain analysis in the
cosmological and astrophysical parameter space by varying all the
parameters that impact on the flux statistics. We have the following
set of cosmological parameters: $H_0$, $n_s$, $\sigma_8$, $\Omega_m$,
a parameter describing the effect of reionization, and the coupling
$\beta$; in addition to these we have the following astrophysical
parameters: $\tau_{\rm eff}$ (amplitude and slope at $z=3$), $\gamma$
(amplitude and two slopes at $z<3$ and $z>3$, $T_0$ (amplitude and two
slopes at $z<3$ and $z>3$), where both the latter functions are parameterized as $A\,[1+z/4]^S$, and a parameter describing the errors
induced by continuum fitting on the flux pdf \citep[see][]{vbh09}.  We
use two data sets: the flux pdf of \cite{tkim} which consists of 63
data points (21 data points per redshift bin) at $z=2.07, 2.52$ and
$z=2.94$) and the 132 data points at $z=2.2,...,4.2$ of the SDSS flux
power of \cite{mcdonald06} (12 measurements of the flux power at
$0.00141<$ k (s/km)$<0.01778$ for the 11 redshift bins). We include
the data covariance matrix in the analysis and present results in
terms of marginalized values for $\beta$.  For the joint analysis of pdf
and flux power we have a total of 15 parameters that are varied and we
apply a weak prior on $\tau_{\rm eff}$ (amplitude and slope). There
are also other 13 parameters that describe noise and resolution
properties and the presence of damped \lya systems for the flux power.

We summarize here the constraints found.  For the pdf: $\beta=0.08\pm
0.05$ and $\beta<0.19$ (2$\sigma$ C.L.)  by using the flux pdf alone in
the range $F=[0.1-0.8]$, with a reduced $\chi^2/\nu=1.09$ (35 d.o.f.);
$\beta=0.04\pm 0.04$ and $\beta<0.1$ (2$\sigma$ C.L.)  by using the
flux pdf alone in the whole range $F=[0-1]$, with a reduced
$\chi^2/\nu=1.21$ (53 d.o.f.). The ranges at low and high
transmissivity are those that are most difficult to model due to the
presence of strong systems and continuum fitting errors,
respectively. Thus, we regard the first result presented as more
conservative even though we do model continuum fitting errors and
correct for numerical resolution \citep{vbh09}.  For the flux power only we
obtain: $\beta=0.07\pm 0.04$ and $\beta<0.14$ (2$\sigma$ C.L.) using
all the 132 data points ($\chi^2/\nu=1.16$, for 120 d.o.f.). All these
numbers are reasonable and demonstrate that the regions of high
transmissivity have a constraining power which is stronger than the
power spectrum alone. If we combine the two measurements we find the
same trends as in \cite{vbh09}: there is not a very good fit to the
data ($\chi^2=200$ for 164 d.o.f.), and a reasonable $\chi^2$ is obtained
only when neglecting the three highest redshift bins of the SDSS flux
power. In this case, we obtain $\beta=0.05\pm 0.03$ and $\beta<0.1$
(2$\sigma$ C.L.)  with a reduced $\chi^2/\nu=1.09$ (146 d.o.f.). All
the other parameters are not affected significantly by the new
parameter introduced and there are not strong degeneracies for
$\beta$.

From the analysis performed we can conclude that robust
 $2\sigma$ upper limits on the coupling constant are in the
range $\beta<0.1-0.2$ (depending on the subset of data chosen).  These
bounds are exclusively derived by the analysis of the observed
properties of the IGM and represent a completely new and independent
test of cDE cosmologies w.r.t. previous constraints \citep[as \eg
][]{Bean_etal_2008,LaVacca_etal_2009,xia09}.
We regard a $2\sigma $ limit of $\beta \lesssim 0.15$ as
a conservative overall bound once the statistical limitations of the different 
samples are taken into account.

\section{Conclusions}

In this work we have explored the possibility of constraining the
coupling $\beta $ between CDM and DE through the statistical
properties of the transmitted flux in \lya forest QSO spectra at
$z=2-4.2$. For this purpose, we have performed the first high-resolution
hydrodynamical simulations  with gas cooling and star formation
in the context of cDE models and quantitatively exploited the capabilities
of flux 1-pt and 2-pt functions to constrain the strength of the coupling
$\beta$ between DE and CDM.

The main results can be summarized as follows:\begin{itemize}
\item
A non-zero coupling between dark matter and dark energy produces
differeces in the probability distribution functions for the gas that are more
prominent in underdense regions: the pdf being skewed towards voids in
cDE models compared to standard cosmologies;

\item Voids in cDE cosmologies are emptier and contain less neutral hydrogen as 
compared to $\Lambda $CDM, and this effect might alleviate tensions between
simulations and observations in voids;

\item
The matter power spectra, for the specific cDE models  investigated here, are also affected
in a scale-dependent way at high redshift to a level that can be constrained 
by the data;
\item
The impact of these effects in terms of flux pdf and power spectra is
smaller but still larger than present statistical errors and more
importantly is not degenerate with that of other parameters that act
in a similar way due to its different redshift evolution;
\item
By using a Monte Carlo Markov Chain
scheme that allows to vary all the parameters involved (astrophysical,
cosmological and noise-related) we obtained a robust and conservative
upper limit of $\beta \lesssim 0.15$ at the $2\sigma$ C.L., after having marginalized over
the other parameters.  This limit on the coupling $\beta $ is a new and completely independent
constraint with respect to previous bounds based on different observables.
\end{itemize}

This work quantitatively shows that the \lya range of scales and
redshifts, where the growth of structures can be radically different
from that measured from a naive extrapolation of either local or very
high redshift probes, is promising for constraining coupled dark
energy cosmologies. The increasing number of QSO spectra that are
being collected (e.g.~BOSS\footnote{http://www.sdss3.org/},
X-Shooter\footnote{http://www.eso.org/sci/facilities/paranal/instruments/xshooter/})
offers the exciting prospect of further improving the numbers and
of understanding in a more refined way, by performing simulations and by
addressing systematic errors, the impact that coupled dark energy
cosmologies can have on the diffuse gas at high redshift.

\section*{Acknowledgments.}
  MB is supported by the DFG Cluster of Excellence ``Origin and
  Structure of the Universe" and partly supported by the TRR
  Transregio Collaborative Research Network on the ``Dark
  Universe". MV is partly supported by ASI/AAE, INFN-PD51 and
  PRIN/MIUR. Numerical simulations have been performed at RZG Computing Centre in Garching.
  Post processing and data analysis have been carried out at COSMOS and
  HPCS (Cambridge), and CINECA thanks to a CINECA/INAF
  grant.  \bibliographystyle{mn2e} \bibliography{master2.bib}

\newcommand{\noopsort}[1]{}
\begin{thebibliography}{}

\bibitem[\protect\citeauthoryear{Amendola}{Amendola}{2000}]{Amendola_2000}
Amendola L.,  2000, Phys. Rev., D62, 043511

\bibitem[\protect\citeauthoryear{Armendariz-Picon, Mukhanov \&
  Steinhardt}{Armendariz-Picon et~al.}{2000}]{ArmendarizPicon_etal_2000}
Armendariz-Picon C.,  Mukhanov V.~F.,    Steinhardt P.~J.,  2000, Phys. Rev.
  Lett., 85, 4438

\bibitem[\protect\citeauthoryear{Baldi}{Baldi}{2010}]{Baldi_2010}
Baldi M.,  2010, arXiv:1005.2188

\bibitem[\protect\citeauthoryear{Baldi \& Pettorino}{Baldi \&
  Pettorino}{2010}]{Baldi_Pettorino_2010}
Baldi M.,  Pettorino V.,  2010, arXiv:1006.3761

\bibitem[\protect\citeauthoryear{{Baldi}, {Pettorino}, {Robbers} \&
  {Springel}}{{Baldi} et~al.}{2010}]{Baldi_etal_2010}
{Baldi} M.,  {Pettorino} V.,  {Robbers} G.,    {Springel} V.,  2010, \mnras,
  403, 1684

\bibitem[\protect\citeauthoryear{Bean, Flanagan, Laszlo \& Trodden}{Bean
  et~al.}{2008}]{Bean_etal_2008}
Bean R.,  Flanagan E.~E.,  Laszlo I.,    Trodden M.,  2008, Phys. Rev., D78,
  123514

\bibitem[\protect\citeauthoryear{{Bi} \& {Davidsen}}{{Bi} \&
  {Davidsen}}{1997}]{bi}
{Bi} H.,  {Davidsen} A.~F.,  1997, \apj, 479, 523

\bibitem[\protect\citeauthoryear{{Bolton}, {Viel}, {Kim}, {Haehnelt} \&
  {Carswell}}{{Bolton} et~al.}{2008}]{bolton07}
{Bolton} J.~S.,  {Viel} M.,  {Kim} T.-S.,  {Haehnelt} M.~G.,    {Carswell}
  R.~F.,  2008, \mnras, 386, 1131

\bibitem[\protect\citeauthoryear{{Croft}, {Weinberg}, {Bolte}, {Burles},
  {Hernquist}, {Katz}, {Kirkman} \& {Tytler}}{{Croft} et~al.}{2002}]{croft02}
{Croft} R.~A.~C.,  {Weinberg} D.~H.,  {Bolte} M.,  {Burles} S.,  {Hernquist}
  L.,  {Katz} N.,  {Kirkman} D.,    {Tytler} D.,  2002, \apj, 581, 20

\bibitem[\protect\citeauthoryear{Di~Porto \& Amendola}{Di~Porto \&
  Amendola}{2008}]{DiPorto_Amendola_2008}
Di~Porto C.,  Amendola L.,  2008, Phys. Rev., D77, 083508

\bibitem[\protect\citeauthoryear{Farrar \& Peebles}{Farrar \&
  Peebles}{2004}]{Farrar_Peebles_2004}
Farrar G.~R.,  Peebles P. J.~E.,  2004, Astrophys. J., 604, 1

\bibitem[\protect\citeauthoryear{Honorez, Reid, Mena, Verde \& Jimenez}{Honorez
  et~al.}{2010}]{Honorez_etal_2010}
Honorez L.~L.,  Reid B.~A.,  Mena O.,  Verde L.,    Jimenez R.,  2010,
  arXiv:1006.0877

\bibitem[\protect\citeauthoryear{Kesden \& Kamionkowski}{Kesden \&
  Kamionkowski}{2006}]{Kesden_Kamionkowski_2006}
Kesden M.,  Kamionkowski M.,  2006, Phys. Rev., D74, 083007

\bibitem[\protect\citeauthoryear{{Keselman}, {Nusser} \& {Peebles}}{{Keselman}
  et~al.}{2010}]{Keselman10}
{Keselman} J.~A.,  {Nusser} A.,    {Peebles} P.~J.~E.,  2010, \prd, 81, 063521

\bibitem[\protect\citeauthoryear{{Kim}, {Bolton}, {Viel}, {Haehnelt} \&
  {Carswell}}{{Kim} et~al.}{2007}]{tkim}
{Kim} T.~.,  {Bolton} J.~S.,  {Viel} M.,  {Haehnelt} M.~G.,    {Carswell}
  R.~F.,  2007, \mnras, 382, 1657

\bibitem[\protect\citeauthoryear{Komatsu et~al.,}{Komatsu
  et~al.}{2009}]{wmap5}
Komatsu E.,  et~al., 2009, Astrophys. J. Suppl., 180, 330

\bibitem[\protect\citeauthoryear{La~Vacca, Kristiansen, Colombo, Mainini \&
  Bonometto}{La~Vacca et~al.}{2009}]{LaVacca_etal_2009}
La~Vacca G.,  Kristiansen J.~R.,  Colombo L. P.~L.,  Mainini R.,    Bonometto
  S.~A.,  2009, JCAP, 0904, 007

\bibitem[\protect\citeauthoryear{Li \& Barrow}{Li \&
  Barrow}{2010}]{Li_Barrow_2010}
Li B.,  Barrow J.~D.,  2010, 1005.4231

\bibitem[\protect\citeauthoryear{{Lidz}, {Faucher-Giguere}, {Dall'Aglio},
  {McQuinn}, {Fechner}, {Zaldarriaga}, {Hernquist} \& {Dutta}}{{Lidz}
  et~al.}{2009}]{lidz09}
{Lidz} A.,  {Faucher-Giguere} C.,  {Dall'Aglio} A.,  {McQuinn} M.,  {Fechner}
  C.,  {Zaldarriaga} M.,  {Hernquist} L.,    {Dutta} S.,  2009, ArXiv e-prints

\bibitem[\protect\citeauthoryear{Macci\`{o}, Quercellini, Mainini, Amendola \&
  Bonometto}{Macci\`{o} et~al.}{2004}]{Maccio_etal_2004}
Macci\`{o} A.~V.,  Quercellini C.,  Mainini R.,  Amendola L.,    Bonometto
  S.~A.,  2004, Phys. Rev., D69, 123516

\bibitem[\protect\citeauthoryear{{McDonald}, {Seljak}, {Cen}, {Shih},
  {Weinberg}, {Burles}, {Schneider}, {Schlegel}, {Bahcall}, {Briggs},
  {Brinkmann}, {Fukugita}, {Ivezi{\'c}}, {Kent} \& {Vanden Berk}}{{McDonald}
  et~al.}{2005}]{mcdonald05}
{McDonald} P.,  {Seljak} U.,  {Cen} R.,  {Shih} D.,  {Weinberg} D.~H.,
  {Burles} S.,  {Schneider} D.~P.,  {Schlegel} D.~J.,  {Bahcall} N.~A.,
  {Briggs} J.~W.,  {Brinkmann} J.,  {Fukugita} M.,  {Ivezi{\'c}} {\v Z}.,
  {Kent} S.,    {Vanden Berk} D.~E.,  2005, \apj, 635, 761

\bibitem[\protect\citeauthoryear{{McDonald}}{{McDonald}}{2006}]{mcdonald06}
{McDonald} P. e.~a.,  2006, \apjs, 163, 80

\bibitem[\protect\citeauthoryear{{Meiksin}}{{Meiksin}}{2009}]{meiksin09}
{Meiksin} A.~A.,  2009, Reviews of Modern Physics, 81, 1405

\bibitem[\protect\citeauthoryear{Nusser, Gubser \& Peebles}{Nusser
  et~al.}{2005}]{Nusser_Gubser_Peebles_2005}
Nusser A.,  Gubser S.~S.,    Peebles P. J.~E.,  2005, Phys. Rev., D71, 083505

\bibitem[\protect\citeauthoryear{{Peebles}}{{Peebles}}{2001}]{Peebles_01}
{Peebles} P.~J.~E.,  2001, \apj, 557, 495

\bibitem[\protect\citeauthoryear{Pettorino \& Baccigalupi}{Pettorino \&
  Baccigalupi}{2008}]{Pettorino_Baccigalupi_2008}
Pettorino V.,  Baccigalupi C.,  2008, Phys. Rev., D77, 103003

\bibitem[\protect\citeauthoryear{Ratra \& Peebles}{Ratra \&
  Peebles}{1988}]{Ratra_Peebles_1988}
Ratra B.,  Peebles P. J.~E.,  1988, Phys. Rev., D37, 3406

\bibitem[\protect\citeauthoryear{Schmidt}{Schmidt}{2009}]{Schmidt_2009}
Schmidt F.,  2009, Phys. Rev., D80, 043001

\bibitem[\protect\citeauthoryear{{Seljak}}{{Seljak}}{2005}]{seljak06}
{Seljak} U. e.~a.,  2005, \prd, 71, 103515

\bibitem[\protect\citeauthoryear{{Springel}}{{Springel}}{2005}]{springel}
{Springel} V.,  2005, \mnras, 364, 1105

\bibitem[\protect\citeauthoryear{{Viel}, {Becker}, {Bolton}, {Haehnelt},
  {Rauch} \& {Sargent}}{{Viel} et~al.}{2007}]{viel08}
{Viel} M.,  {Becker} G.~D.,  {Bolton} J.~S.,  {Haehnelt} M.~G.,  {Rauch} M.,
  {Sargent} W.~L.~W.,  2007, ArXiv e-prints, 709

\bibitem[\protect\citeauthoryear{{Viel}, {Bolton} \& {Haehnelt}}{{Viel}
  et~al.}{2009}]{vbh09}
{Viel} M.,  {Bolton} J.~S.,    {Haehnelt} M.~G.,  2009, \mnras, 399, L39

\bibitem[\protect\citeauthoryear{{Viel}, {Branchini}, {Dolag}, {Grossi},
  {Matarrese} \& {Moscardini}}{{Viel} et~al.}{2009}]{vielng}
{Viel} M.,  {Branchini} E.,  {Dolag} K.,  {Grossi} M.,  {Matarrese} S.,
  {Moscardini} L.,  2009, \mnras, 393, 774

\bibitem[\protect\citeauthoryear{{Viel} \& {Haehnelt}}{{Viel} \&
  {Haehnelt}}{2006}]{vielhaehnelt06}
{Viel} M.,  {Haehnelt} M.~G.,  2006, \mnras, 365, 231

\bibitem[\protect\citeauthoryear{{Viel}, {Haehnelt} \& {Springel}}{{Viel}
  et~al.}{2004}]{viel04}
{Viel} M.,  {Haehnelt} M.~G.,    {Springel} V.,  2004, \mnras, 354, 684

\bibitem[\protect\citeauthoryear{Wetterich}{Wetterich}{1988}]{Wetterich_1988}
Wetterich C.,  1988, Nucl. Phys., B302, 668

\bibitem[\protect\citeauthoryear{Wetterich}{Wetterich}{1995}]{Wetterich_1995}
Wetterich C.,  1995, Astron. Astrophys., 301, 321

\bibitem[\protect\citeauthoryear{Wintergerst \& Pettorino}{Wintergerst \&
  Pettorino}{2010}]{Wintergerst_Pettorino_2010}
Wintergerst N.,  Pettorino V.,  2010, 1005.1278

\bibitem[\protect\citeauthoryear{{Xia}}{{Xia}}{2009}]{xia09}
{Xia} J.,  2009, \prd, 80, 103514

\end{thebibliography}

\end{document}